\documentclass[%
 reprint,
 amsmath,amssymb,
 aps,
]{revtex4-1}

\usepackage{graphicx}
\usepackage{dcolumn}
\usepackage{bm}

\begin{document}

\preprint{APS/123-QED}

\title{Interface of graphene nanopore and hexagonal boron nitride as a sensing device}

\author{Fabio A. L. de Souza}
 \altaffiliation[Also at ]{Division of Materials Theory, Department of Physics and Astronomy, Uppsala University, Box-516, SE 75120, Sweden.}
\author{ W.L Scopel}%
\affiliation{%
 Departamento de F\'{\i}sica, Universidade Federal do Esp\'{\i}rito Santo- UFES, Vit\'oria/ES, Brazil.
}%


\author{Rodrigo G. Amorim}
\altaffiliation[Also at ]{Division of Materials Theory, Department of Physics and Astronomy, Uppsala University, Box-516, SE 75120, Sweden.}
\email{rgamorim@id.uff.br}
\affiliation{Departamento de Ci\^encias Exatas, Universidade Federal Fluminense-UFF, Volta Redonda, RJ, Brazil.
}%
\author{Ralph H. Scheicher}
\affiliation{%
Division of Materials Theory, Department of Physics and Astronomy, Uppsala University, Box-516, SE 75120, Sweden.
}%


\date{\today}

\begin{abstract}
The atomically-precise controlled synthesis of graphene stripes embedded in hexagonal boron nitride opens up new possibilities for the construction of nanodevices with applications in sensing. Here, we explore properties related to electronic structure and quantum transport of a graphene nanoroad embedded in hexagonal boron nitride, using a combination of density functional theory and the non-equilibrium Green's functions method to calculate the electric conductance. We find that the graphene nanoribbon signature is preserved in the transmission spectra and that the local current is mainly confined to the graphene domain. When a properly sized nanopore is created in the graphene part of the system, the electronic current becomes restricted to a carbon chain running along the border with hexagonal boron nitride. This circumstance could allow the hypothetical nanodevice to become highly sensitive to the electronic nature of molecules passing through the nanopore, thus opening up ways for the detection of gas molecules, amino acids, or even DNA sequences based on a measurement of the real-time conductance modulation in the graphene nanoroad.
\end{abstract}

\pacs{Valid PACS appear here}
\maketitle


\section{Introduction}

Two-dimensional materials have gained enormous attention recently owing to their novel properties and potential applications in future generations of nanodevices\cite{min2011fast,Lee2013,Prasongkit2011,Amorim2015}. Among them, graphene and single-layer hexagonal boron nitride (h-BN) share similar structural characteristic, such as a comparable lattice constant with a small mismatch (less than 2\%), while possessing drastically different electronic properties. Graphene\cite{Novoselov2004,Novoselov2005} is a zero-gap semiconductor with high carrier mobility at room temperature, whereas a monolayer of h-BN\cite{Kubota2007,Song2010} is a semiconductor with a wide direct bandgap of around 4.6$-$6.0 eV\cite{blase1995quasiparticle,Amorim2013}. The comparable structures of graphene and h-BN offers the possibility to have both materials joined together in a single layer, forming a hybrid 2D material. 

Experimental hybrid structures have been synthesized, in particular different shapes (triangular/hexagonal) of h-BN embedded in graphene\cite{expIV,Liu2014}, BN tube growth on top of graphene\cite{parashar2015switching} as well as graphene stripes entrenched in h-BN\cite{Liu2013,Levendorf2012}. In the latter example, the nanoroads were created using chemical vapour deposition (CVD), where firstly an h-BN monolayer was synthesized and then partially etched through exposure to laser-cut masks; secondly, a few layers of graphene were grown on the etching region at high temperature ($>900^\circ$C) and using CH$_4$ as a carbon source. In this manner, the size and shape of the graphene domains can be controlled very precisely.

A different type of nano-sized structures relevant for the present work are nanopores. Their controlled fabrication in solid-state materials still poses some challenges. In recent studies\cite{VandenHout2010}, nanopores were created on SiN membranes and it was claimed that it is possible to control the stability, size and also the shape of the nanopores, using transmission electron microscope (TEM). Another technique, using electrochemical reaction (ECR) \cite{feng2015electrochemical}, can create a nanopore in 2D materials with ann atom-by-atom control, and this technique has been applied on both graphene and MoS$_2$. 

\begin{figure*}[ht!]
\centering
  \includegraphics[height=8.cm]{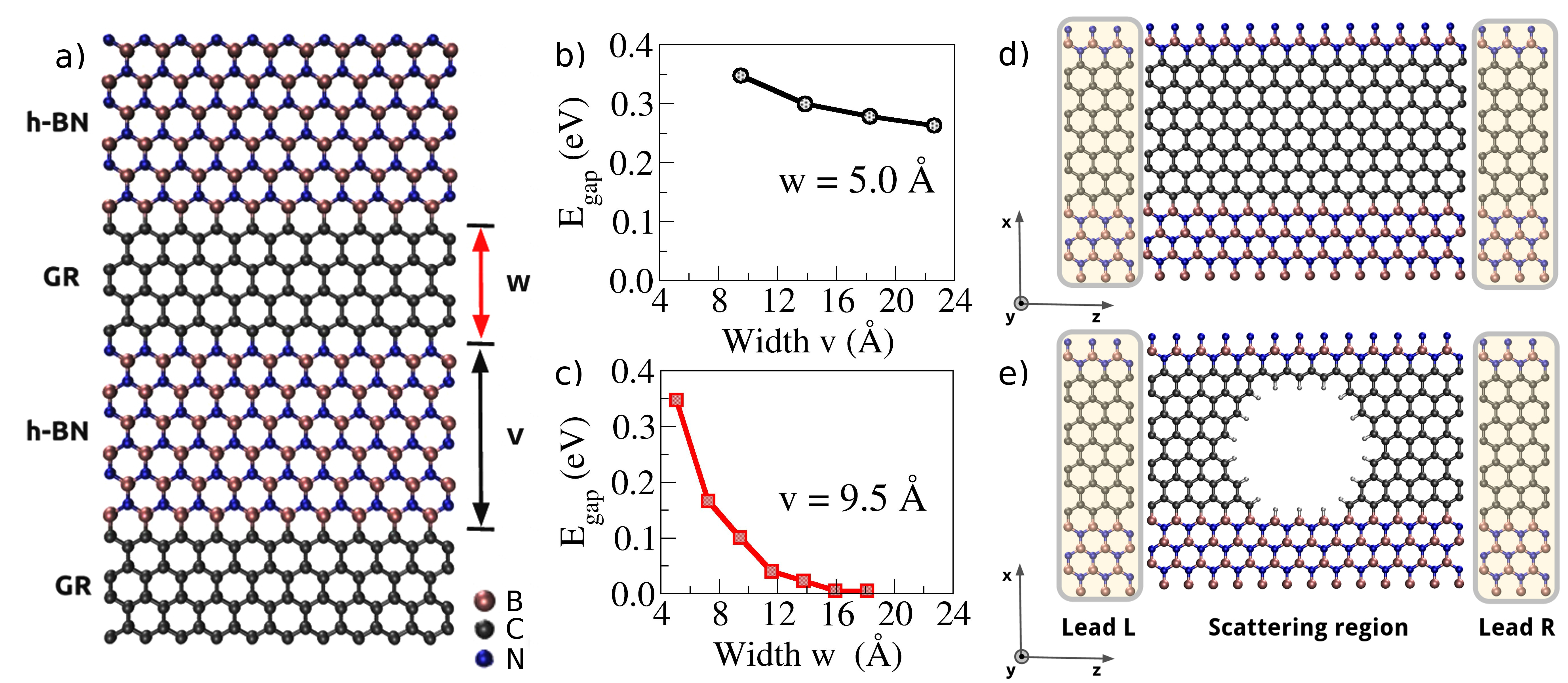}
  \caption{ (a) The relaxed structure of the hybrid system composed of graphene and h-BN stripes with a zigzag interface, where $w$ and $v$ indicate the width of graphene and h-BN, respectively. (b) The behavior of the bandgap as a function of width $v$ for fixed $w$, and (c) the behavior of the bandgap as a function of width $w$ for fixed $v$. (d) The sketch of the model used to calculate electronic transport properties of a graphene nanoroad embedded in h-BN, and (e) the sketch of the model used to calculate electronic transport properties of a graphene nanoroad including a nanopore embedded in h-BN.}
  \label{structure}
\end{figure*}

Inspired by these experimental development, we explore here a setup which combines a hybrid structure composed of a graphene stripe (or nanoroad) embedded in h-BN, and a nanopore in the graphene part. A possible fabrication process for such an architecture could be the above mentioned electrochemical reaction process, starting out from a defect in graphene and growing until reaching the carbon atoms bordering with h-BN on either side of the nanoroad. 

We have explored how the electronic properties of such a hybrid system depend on the graphene stripe width and predict quantum transport properties for both the pristine nanoroad system as well as for the system containing a nanopore sized in such a way as to border on a one-atom wide carbon chain at one edge which is confined by h-BN on the other side. 

It is found by us that for a specific range of energies at which electrons are injected into this hypothetical nanodevice, it would be possible to control the local current path. For the pristine system, the current can run through either the borders or the middle of the device; for the system containing a nanopore, the current is confined to the aforementioned carbon chain.

\section{Methodology}

Figure \ref{structure}-a shows schematically the hybrid graphene/h-BN system similar to the experimental nanoroads synthesized previously by Liu et al. \cite{expIV}. In this setup, a graphene domain is embedded in a h-BN sheet and the system can be characterized by two width parameters ($w$ and $v$). In this graphene/h-BN system, we notice two different interfaces, one with C-B and the other with C-N bonds. 

For modeling, we employed \textit{ab initio} Density Functional Theory (DFT) \cite{hohenberg1964inhomogeneous,kohn1965self} as implemented in the SIESTA \cite{soler2002siesta} code. The exchange-correlation potential was treated with the Generalized Gradient Approximation of Perdew-Burke-Ernzerhof (GGA-PBE)\cite{perdew1996generalized}. Core electrons were modeled with Troullier-Martins norm-conserving pseudopotentials\cite{troullier1991efficient} and the valence electrons functions were expanded in a double-zeta polarized basis set (DZP) of localized orbitals, and the real space grid (energy cutoff) was set to be 200 Ry. Structural relaxations were performed ensuring that the Hellmann-Feynman forces acting on each ion are less than 0.01 eV/\AA. A Monkhorst-Pack scheme with a mesh of 3$\times$1$\times$2 $k$-points for the Brillouin zone integration was used for structural relaxation, while 52$\times$1$\times$78 $k$-points were used for density of states (DOS) calculations. The relaxed lattice constant obtained for the hybrid system was 2.52 \AA, and the C-C, B-N, B-C and C-N bond lengths are 1.45, 1.46, 1.56 and 1.41 \AA, respectively. Two setups were investigated as shown in Figure \ref{structure} d-e with equal supercell size. For both systems (hybrid and pore), the scattering region measures 26.2 {\AA} in $x$-direction and 32.8 {\AA} in $z$-direction.
For transport calculation we used a combination of DFT with the non-equlibrium Green's functions method as implemented in TranSiesta\cite{brandbyge2002density}. Figure \ref{structure} d-e shows the transport setup composed of two electrodes (left and right) and a scattering region in between. The leads act as electron reservoirs and are semi-infinite. To solve this non-periodic problem it is possible to write the Green's functions as
\begin{equation}
{\cal G}\left(E,V\right) = \left[ E\times S_{\mathrm S} - H_{\mathrm S}\left[\rho\right] - \Sigma_{\mathrm L}\left(E,V\right) - \Sigma_{\mathrm R}\left(E,V\right) \right]^{-1} ~,
\end{equation}
where $S_S$ and $H_S$ are the overlap matrix and Hamiltonian for the scattering region, respectively. The effects of both semi-infinite leads are given by the respective self-energies $\Sigma_{L/R}$. In transport simulations, the charge density is calculated via Green's functions self-consistently, and the initial guess for the charge density is obtained from a DFT calculation. When convergence is achieved, the transmission coefficient can be calculated as
\begin{equation}
T\left(E \right) = Tr \left[\Gamma_{\mathrm L}\left(E,V\right) {\cal G}\left(E,V\right) \Gamma_{\mathrm R}\left(E,V\right) {\cal G}^{\dagger}\left(E,V\right) \right]
\end{equation}
where the coupling matrices are given by $\Gamma_\alpha = i \left[ \Sigma_\alpha - \Sigma_\alpha^{\dagger} \right]$, with $\alpha\equiv \left\{{\mathrm{L,R}}\right\}$. The physical meaning for transmittance is the probability of an incoming electron with energy $E$ entering the scattering region from the left electrode to reach the right electrode. Finally, we employ the general definition for electric current using the Keldysh formalism, and obtain the current density between two sites $M$ and $N$ as
\begin{equation}
i\left(E\right)_{N\rightarrow M} = 4\frac{e}{h} \sum_{
\begin{array}{c}
n \in N\\ 
m\in M
\end{array}}
\Im
  \left[ \left\{ {\cal G}\left(E\right) \Gamma_{\mathrm L}{\cal G}^\dagger\left(E\right)\right\}_{mn}  H_{nm} \right]
\end{equation}
where the sum is performed over all localized atomic orbitals $n$ and $m$ of the basis set, which are associated with sites $N$ and $M$, respectively. This quantity is called the ``local current'' and for zero bias we have the transmittance projection between two sites. For more details regarding the theory of electronic transport, we refer to the literature \cite{brandbyge2002density}. 

\begin{figure*}[ht!]
 \centering
 \includegraphics[height=6.5cm]{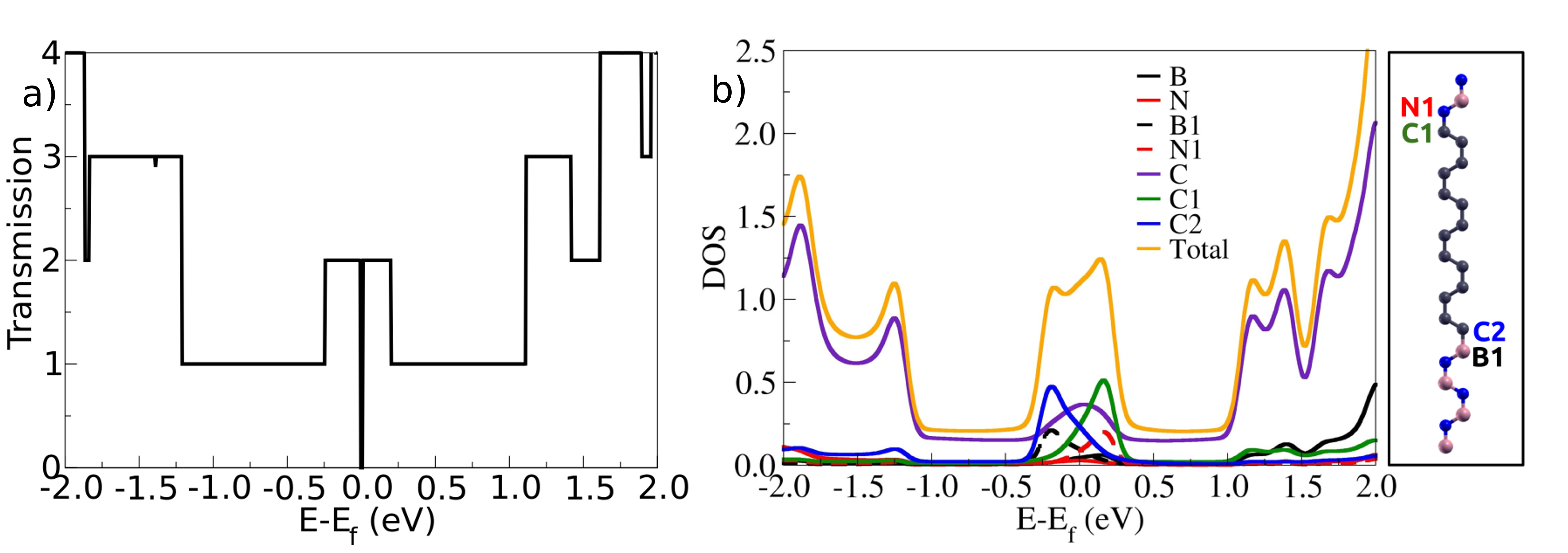}
 \caption{Electronic properties for the pristine graphene/h-BN hybrid system: (a) Zero bias Transmission spectra; (b) Projected Density of States (PDOS). The N1 and C1 labels were used for nitrogen and carbon atoms from the C-N  graphene/BN interface, and B1 and C2 labels were used for boron and carbon atoms from the C-B graphene/BN interface. The contribution of all carbons, except C1 and C2, is given by C, and the total DOS is represented as well.}
 \label{transmission}
\end{figure*}

\section{Results}


 
Figure \ref{structure}-a shows the fully relaxed most stable hybrid system with two interfaces (B-C and C-N). Due to a small mismatch between the lattice parameters of graphene and h-BN (less than 2 $\%$), no buckling is observed in the interface and C-C, C-B, and C-N bond lengths are in good agreement with previous results reported in the literature\cite{bhowmick2011quantum,nguyen2014reactivity}. In order to study the possibility of tuning the bandgap by changing the width of both graphene and h-BN stripe widths two different cases were considered: Figure \ref{structure}b shows the bandgap behavior first for increasing the h-BN stripe width while keeping the width of graphene fixed to w = 5.0 \AA. We note that the bandgap is slightly decreasing by less than 90 meV. When keeping the h-BN widht fixed to v = 9.5 \AA and varying the graphene width, the bandgap can be tuned all the way close to zero (see  Figure \ref{structure}c), leading to a quasi-metallic system. The remaining tiny bandgap is smaller than thermal energy fluctuations at room's temperature. Such bandgap trends are in agreement with previous reported results for hybrid monolayers\cite{jungthawan2011electronic} and nanoribbons\cite{sun2015lateral} in graphene and h-BN. These last findings lead to the hybrid system as proposed in Figure\ref{structure}-d with w and v equal to 13.7 and 9.5 \AA, respectively.
 
Figure \ref{transmission}-a,b shows the calculated zero bias transmission spectra and the PDOS for the hybrid system, respectively. We note one transmission channel in two energy range regions. To better understand contribution from the different atomic species to these plateaus, we analyzed the PDOS for the same energy range. One can see that the major contribution comes from carbon atoms (C) of the central region. The transmission results show the typical signature of zigzag nanoribbon for this energy range, as expected\cite{chauhan2014electronic,liping2011first}.
In the central region there are two transmission channels, with symmetric plateaus around the Fermi level. The transmission enhancement could be associated with interface states, as can be observed in the PDOS (Figure \ref{transmission}-b). More specifically, the left plateau has a dominant contribution from hybridized orbitals from C2 and B1 atoms. However, we also note a non-negligible contribution from C atoms. The right plateau is ascribed to hybridized orbitals from C1, N1 and C atoms. In addition, the transmission probability drops to zero at the Fermi level due to the tiny bandgap (24 meV) in the band structure of the hybrid system.  

\begin{figure}[h!]
 \centering
 \includegraphics[width=9cm]{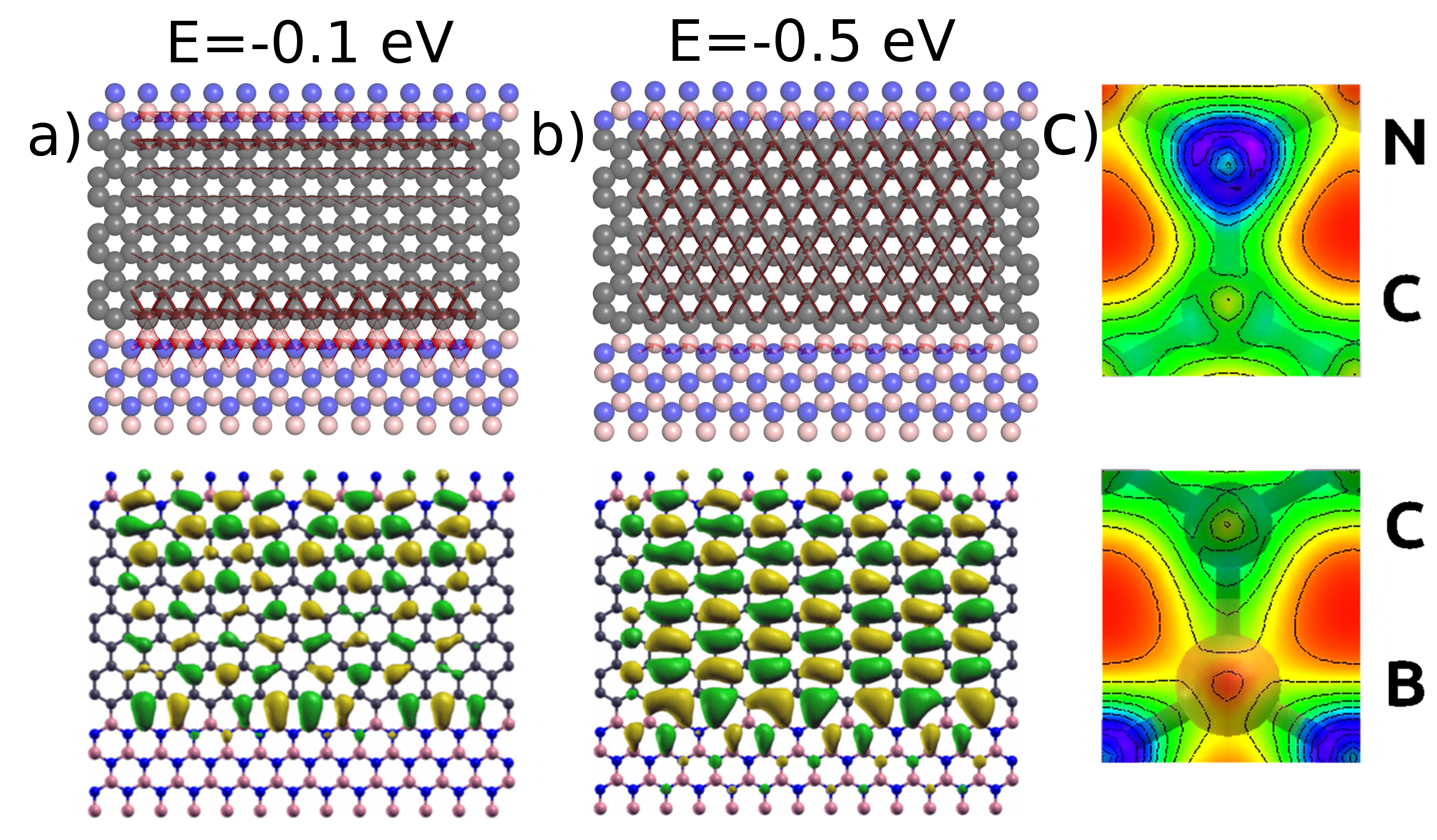}
 \caption{Properties related to electric current and charge density for the hybrid system: Local currents (top panels) and the isosurfaces of the eigenchannel wave functions (bottom panels) colored according to sign, yellow (green) for positive (negative), at two different energies: (a) $-$0.1 eV and (b) $-$0.5 eV. (c) Charge distribution between two atoms at the C-N (top panel) and C-B (bottom panel) interfaces.}
 \label{current}
\end{figure}

\begin{figure*}[ht!]
 \centering
 \includegraphics[height=6cm]{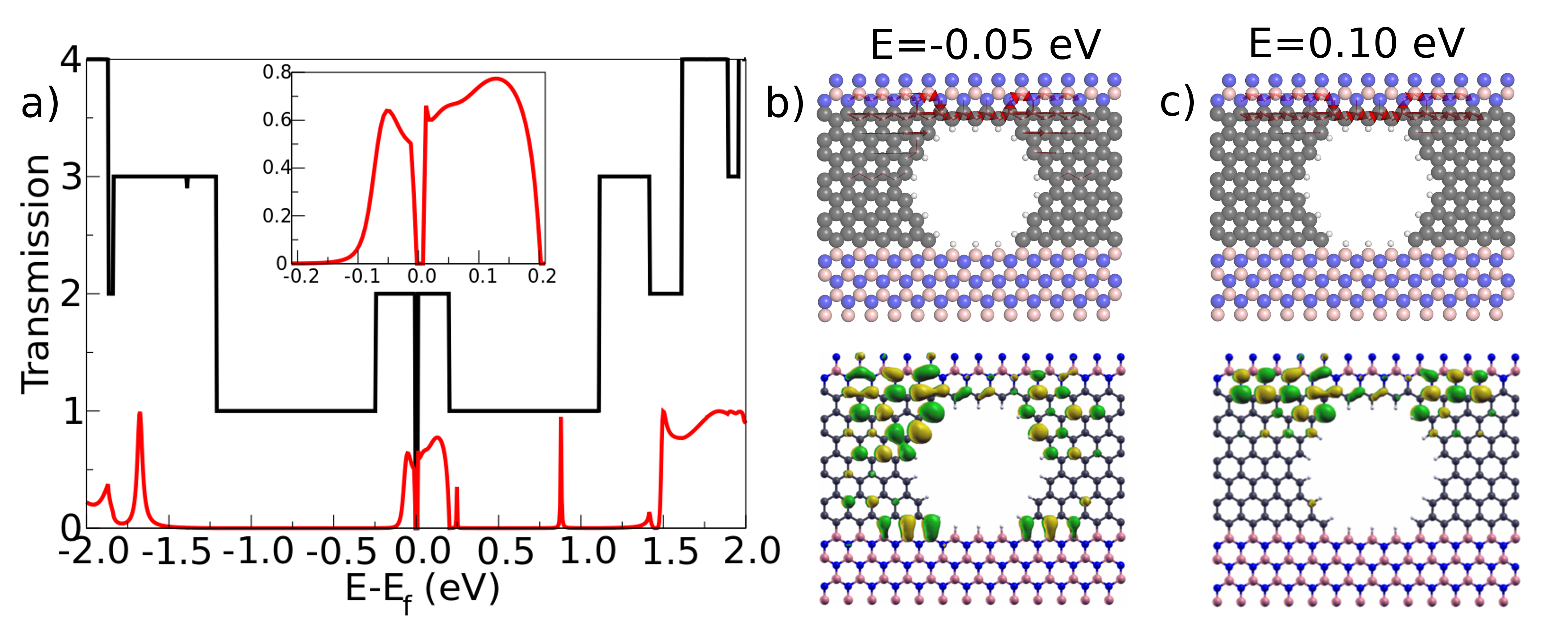}
 \caption{(a) Zero-bias transmission spectra for the hybrid (black line) and pore device (red line). In (b) and (c) the local currents for pore device (top panel) and the wave functions (bottom panel), colored according to sign, yellow for positive and green for negative, at energies E = $-$0.05 eV and E = +0.10 eV.}
 \label{fig:pore}
\end{figure*}

The influence of interface states on the transmission spectra can also be verified through the local currents. However, in this case, the local current means zero-bias transmittance projection between two states, because there are no actual currents involved. Figure \ref{current}-a (top panel) shows, at the energy E = $-$0.1 eV, the local current for the hybrid system. We observe in both interfaces the current accumulation due to interface states. Nevertheless, we can see the biggest current intensity for C-B border due to the high B1-C2 states hybridization at this specific energy. To corroborate this assumption, we analyzed the wave function (WF) for the same energy (Figure \ref{current}-a - bottom panel). The WF is more localized at the borders, and also we noted a distinct behavior for the interface: C-B (WF bonding) and C-N (WF anti-bonding nature). For the central part of the graphene nanoroad, the WF is less localized, in agreement with the local current results. On the other hand, Figure \ref{current}-b (top panel) presents a similar analysis for another energy (E = -0.5 eV). In this case, we can see the projected currents flowing uniformly through the graphene nanoroad central region. The WF spreads uniformly in the graphene nanoroad (Figure \ref{current}-b - bottom panel), again in agreement with the local current trends.

Next we investigated the same system, but with a nanopore cut out in the graphene part with a diameter of 12.5 {\AA}. The resulting dangling bonds were passivated by H atoms. This new system is referred to as ``pore devic''e (Figure \ref{structure}-e). Experimentally, recent progress with a new technique called electrochemical reaction (ECR)\cite{feng2015electrochemical} allows improved control of the nanopore fabrication process. In this novel route to generate nanopores in 2D materials, tested so far only for MoS$_2$ and graphene, the authors suggest that the nanopore growth is started at a single vacancy, the so-called active site (AS), and successive atoms are removed from the neighborhood. Taking in account these issues, we are looking for a region of the hybrid system, where it is most likely for vacancy defects to exist, attempting to get some insight on the AS position. Initially, we will analyze the border atoms as a good candidates for an AS. Figure \ref{current}-c shows the total charge density and it is seen that the bonding for C-N borders is stronger compared to C-B.We explored vacancy creation in two regions: i) h-BN and ii) graphene stripe. For the h-BN case, the formation energies of single boron vacancy at C-B interface and in the middle of the BN stripe are 14.20 and 14.56 eV, respectively. In addition, for the nitrogen defect, formation energies are 8.33 and 9.04 eV, for C-N interface and middle, respectively. However, for the graphene stripe we investigated formation energies for a single carbon vacancy in the middle, at C-N and at C-B interface, where we found 8.40, 6.43, and 6.18 eV, respectively. Thus, based on the formation energies and also the charge density analysis, we might infer that the most probable region for a vacancy defect to exist is in one of the carbon atoms from the graphene stripe at the C-B interface. The carbon atom removal from C-B interface leads to two atoms with dangling bonds in its neighborhood. As a result, it is expected that the energy necessary to remove a second atom in this region decreases, which favor an increase of the void size. Then, considering the hybrid structure containing a single vacancy of carbon at the C-B interface, a second atom should be removed in the neighborhood of the first defect: (i) a boron atom with a dangling bond; (ii) a carbon atom with a dangling bond. The formation energies of these second vacancies are 9.72 and 1.06 eV, respectively. Thus, these results strongly suggest that the increase of the void size might be towards the graphene domain. In this manner, the pore could be continued to be grown electrochemically in the graphen stripe, until almost reaching the h-BN interface. Thus, our proposed nanopore setup in a hybrid system, where a chain of carbon atoms is kept close to the C-N interface, could be considered as feasible. It is important to emphasize that theoretical investigations on the dynamical process of pore growth constitute a complex problem and are beyond the scope of the present work.



Figure \ref{fig:pore}-a shows the zero bias transmission spectra for the both systems: hybrid and pore device. We removed the C atoms from the central region and kept just one atomic wire on the C-N interface, and as a consequence, the one channel regions has the transmission suppressed. It is important to stress that the symmetry is broken in the central region. As a result of the C atoms removal from the C-B interface, the peak shrink and the majority contribution is given to the remaining C-B atoms. For the right plateaus, we note just a transmission decreasing compare to hybrid system, and also the signature from C-N atoms are kept. Figure \ref{fig:pore}-b,c shows the local currents for the two peaks previously analyzed. As expected, the local current flows through the carbon wire in both energies. These results are consistent with the WF localization close to C-N interface. However, analyzing more carefully the WF, we note for negative energies (left plateau) the existence of localized contribution due to remaining C-B atoms. One the other hand, as was demonstrated, there is no considerable contribution from C-B atoms for the right plateau, justifying the absence of the WF in the positive energy in C-B interface.  

Our proposed nano-biodevice studied herein can rely on two sensing mechanisms: (i) sensing molecules that adsorb ``on top'' and (ii) sensing molecules that translocate through the nanopore. The former case utilizes Fano resonances signatures which show up in the transmission function for specific biomolecules at different energies \cite{min2011fast,Amorim2015}. However, in this new proposal there are two new features: (a) the current is confined through the nanoroad without borders defects, with a better control compared to a single material such as graphene or silicene, and (b) there are two possible interfaces (B-C) and (C-N) which could play different roles for molecule adsorption due to the local dipole moment at the interfaces. This device is a good candidate for applications in gas sensors as well as biomolecules detection or protein/DNA sequencing. For the latter case, nanopore with a carbon chain close to (C-N) interface, the sensing mechanism would be given by electric current modulation. The nanopore device transmittance was characterized and we expect a potential candidate for biomolecules sequencing and detection, because the molecules passing through the pore could modulate the conductance and thereby the electric current.        

The hybrid device was synthesized experimentally\cite{expIV}, but controlling the interface and also the width is currently still very challenging. There are enormous experimental efforts underway to control the pore size, shape and stability\cite{VandenHout2010,feng2015electrochemical}, and it can be said that nowadays it remains a significant technological challenge to build a sophisticated nano-biosensor.

\section{Conclusions}

We demonstrated here, using density functional theory combined with non-equilibrium Green's function, that a hybrid system made from graphene and h-BN, and containing a nanopore in the graphene part, possesses the capability to act as a potential nanosensor. It was found by us that changing the width of the graphene stripe can suppress the band gap, inducing a semiconductor to zero-gap-semiconductor transition. The density of states confirms that the hybrid graphene/h-BN transmittance is mostly given by the interface atoms for $T=2G_0$ (central plateau), and the central atoms contribute for the $T=1G_0$. The transmittance was projected at two energies ($-$0.1 eV and $-$0.5 eV), and for the first case the current flows mainly along the interfaces. On the other hand, for the latter energy, the current flows through the central atoms of the graphene stripe. We also demonstrated that electrochemical pore creation could be more probable to start out close to the (B-C) interface. This leads us to propose an architecture which contains a nanopore with a carbon chain close at the (C-N) interface. For this setup, the $T(E)$ is suppressed to zero due to the pore in the energy range from $|1.25|$ to $|0.25|$ eV, and there is a noted transmittance decrease close to the Fermi energy. For the remaining transmission peaks we have shown that the current flows through the carbon chain. In summary, the presented results indicate that both the pristine setup as well as the one containing the nanopore possess great potential to function as a candidate for applications in gas sensing devices as well as DNA sequencing.

\section{Acknowledgements}

F.A.L.S. thanks CAPES Foundation (Ministry of Education of
Brazil, Bras\'{\i}lia-DF 70040-020) for a scholarship (Process No. 1640/14-3). Financial support from the Carl Tryggers Stiftelse and the Swedish Research Council (VR, Grant No. 621-2009-3628) is gratefully acknowledged. The Swedish National Infrastructur for Computing (SNIC) and the Uppsala Multidisciplinary Center for Advanced Computational Science (UPPMAX) provided computing time for this project. 


\bibliography{rsc} 
\bibliographystyle{rsc}
\end{document}